# Artificial Intelligence in Everyday Life 2.0: Educating University Students from Different Majors

Maria Kasinidou
Open University of Cyprus
Nicosia, Cyprus
maria.kasinidou@ouc.ac.cy

Styliani Kleanthous
CYENS Centre of Excellence &
Open University of Cyprus
Nicosia, Cyprus
s.kleanthous@cyens.org.cy

Matteo Busso
University of Trento
Trento, Italy
matteo.busso@unitn.it

Marcelo Rodas
University of Trento &
Fondazione Bruno Kessler
Trento, Italy
mrodasbritez@fbk.eu

Jahna Otterbacher
Open University of Cyprus &
CYENS Centre of Excellence
Nicosia, Cyprus
jahna.otterbacher@ouc.ac.cy

Fausto Giunchiglia
University of Trento
Trento, Italy
fausto.giunchiglia@unitn.it

## ABSTRACT
With the surge in data-centric AI and its increasing capabilities, AI applications have become a part of our everyday lives. However, misunderstandings regarding their capabilities, limitations, and associated advantages and disadvantages are widespread. Consequently, in the university setting, there is a crucial need to educate not only computer science majors but also students from various disciplines about AI. In this experience report, we present an overview of an introductory course that we offered to students coming from different majors. Moreover, we discuss the assignments and quizzes of the course, which provided students with a firsthand experience of AI processes and insights into their learning patterns. Additionally, we provide a summary of the course evaluation, as well as students' performance. Finally, we present insights gained from teaching this course and elaborate on our future plans.

## 1 INTRODUCTION
Artificial intelligence (AI) technologies have become embedded into everyday life, to the point where individuals may use them without even realizing it. For instance, we regularly use AI tools like "Siri" to make a call without touching our phones or face recognition to unlock our phones without using our passwords. The integration of AI into everyday life will only increase as new applications are developed for use in our homes, schools, governments, social lives, and workplaces. But despite the progress made, we have also seen how serious the consequences of misunderstanding or failing to question AI decisions can be – leading to issues such as viral misinformation [8], biased systems that disproportionately impact marginalized communities [1], and serious concerns about data privacy. This situation highlights the need to bridge the gap between AI's everyday presence and people's lack of knowledge, so we can clear up misconceptions, reduce fears, and embrace a more informed relationship with the AI that is shaping our future [4].

Several initiatives have considered the design and evaluation of curriculum and courses on Machine Learning (ML) and AI [14], targeting university students in computer science (CS) [6, 22, 24], students in majors other than CS [25], younger students in middle-school, and K-12 students and teachers [15, 17, 18, 21]. A recent review found that although most initiatives focus on CS students, educating K-12 students on AI is also becoming popular [19].

Much effort by the research community, in collaboration with teachers, has been devoted to understanding how AI education at K-12 should be designed [23]. Co-design workshops with teachers have been previously used in an attempt to identify what should be included in an AI curriculum. Sabuncuoglu [21] delivered a 36-week, open-source AI curriculum to both K-12 students and teachers. They also highlighted teachers' need for directions on what should be taught about AI. Kim et al. [9] proposed an AI curriculum for promoting students' AI literacy to elementary school students that included units for AI, ML, Computer Vision, Machine Translation and Self-driving cars. In [10], authors examined whether after attending three AI courses senior secondary students can understand machine learning, and deep learning concepts and be able to discuss related ethical issues. Their findings indicated that students after attending these courses were able to better understand ML and deep learning and their ethical implications.

A growing number of papers have tried to develop curriculum and courses for university students mainly coming from the CS degrees. In [22] they developed AI and ML courses for undergraduate CS students, as well as graduate students in engineering. Their work provided further insights for the design and effective implementation of such courses, accommodating various modes of instruction, including traditional in-person classes, hybrid setups, and fully online environments. Wunderlich et al. [25] discussed the advantages of introducing ML to Business major students. Focusing on how ML is applied to real-world business problems, they argued that education on ML is necessary for the future workforce.

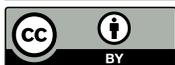

A few initiatives have aimed to provide AI courses to university students coming from diverse educational backgrounds [11, 12]. In [11], authors explored whether students coming from different backgrounds could develop a conceptual understanding of AI after attending a course. They assessed before-and-after-course tests and found that students made significant progress in their understanding of AI. Similarly, in [12], 36 university students with diverse backgrounds completed an AI program focused on ML and deep learning. They achieved a notable improvement in students' understanding of AI concepts, AI literacy, and ethical awareness. However, it is important to note that these courses mainly focused on the technical aspects of AI, rather than delving into what AI is and considering its impact in the broader social context.

Thus, current approaches are primarily based on technical content and aim to teach students how to use machine learning algorithms for developing AI applications. This requires students to have basic programming skills and computational thinking. The reality is that students coming from disciplines other than computer science or engineering-related degrees often lack such skills and knowledge, likely finding it difficult to follow a technical course.

In this experience report, we offer a comprehensive overview of our course and its assignments. Additionally, we elaborate on the learning activities and how they grant students a hands-on experience with AI systems and offer insights into how these systems are developed (Section 2). Moreover, we provide a summary of the evaluation of the course along with the performance of the students (Section 4). Finally, we share our experience of teaching this course and provide suggestions for adopting the course (Section 5).

## 2 COURSE DESIGN

The course was offered as optional from the School of Innovation at the University of Trento, for 2 ECTS [1]. Students received a pass/fail grade. To pass, students needed to attend at least 80% of the course meetings, and complete all the assignments and the final quiz.

### 2.1 Learning Objectives

The learning objectives were displayed on the course website[2], as well as the syllabus, and were discussed during the first meeting with the participants. The main objective of the course was to help students appreciate the nature of today's data-driven AI, as well as to understand the basic elements of AI technologies, and how they are developed. It was expected that students would become familiar with the fundamental concepts around AI, and the widespread methods used in the creation of systems, in particular, ML.

Upon completion, students were also expected to be familiar with key AI technologies, such as Computer Vision (CV), Natural Language Processing (NLP), and Personalization, and be able to recognize them in their daily interactions, as well as to understand the dangers and limitations of such technologies. In addition, they were expected to understand the ethical and social implications that AI technologies bring with them and be able to analyze the functions of "everyday" AI applications and understand the potential risks of these technologies, and be able to critically evaluate them.

### 2.2 Structure

The course was developed in English and consists of nine units – eight learning units and one revision and reflection session. The course delivery followed a blended learning approach, combining pre-recorded lectures and synchronous interactive sessions. Lectures were pre-recorded and made available on the course website. The pre-recorded lectures had a duration between 15 to 25 minutes. Students were expected to watch the pre-recorded lectures before attending the synchronous sessions.

Synchronous sessions – which were recorded and archived on the course's website – took place once per week, and lasted for one hour. They began with a brief recap of the week's lecture, lasting approximately 10 minutes. After the recap, a question and answer (Q&A) session and an interactive activity followed. During the Q&A, students had the opportunity to clarify any question or seek further explanation on the unit's content. Interactive activities were comprised of group discussions, hands-on exercises or case studies, providing participants with an opportunity to actively apply their knowledge and enhance their understanding of each unit.

### 2.3 Learning Units

**Introduction.** The objectives of the first unit are threefold. First, it aimed to familiarize students with the fundamental principles of online and hybrid learning. Given that our course was delivered through our platform, it was crucial for students to gain proficiency in navigating and engaging with the online learning environment. Secondly, this unit provided students with a comprehensive understanding of the objectives of the course.

During the live session, students were presented with the learning units and briefed on the prerequisites for successful completion. Additionally, a brief introduction to AI was provided and a demonstration of the course website was conducted, encompassing guidance on its utilization for assignment and quiz submission. Finally, we introduced students to iLog [2, 26], an AI tool designed to collect and process data, demonstrating how this data can be utilized to draw conclusions.

**Analyzing AI applications.** In the second unit, the concept of intelligence in machines was explored, highlighting that machines are goal-oriented and have rational decision-making abilities within a specific scope. The learning objectives included identifying common characteristics of AI applications, describing their basic components, analyzing their functionalities, evaluating their "intelligent" nature, and examining everyday AI applications to identify their functionalities, data sources, and positive and negative behavioral traits for end users. This unit aimed to provide a comprehensive understanding of AI applications and their impact using examples of applications we use in our daily lives.

During the live session, students were given examples of AI applications like a robotic vacuum, a smart toothbrush and Netflix.

---

[1] European Credit Transfer and Accumulation System (ECTS).
[2] http://www.protea.space/course/ai-in-everyday-life/

The focus was on their "smart" functionalities, including data collection, decision-making, and activation. Then, they were prompted to recognize the "smart" functionalities in other everyday AI applications, like social media newsfeeds. They were then asked to contribute to a shared document, describing the smart behaviors of social media newsfeeds and their functions.

**Machine Learning and Big Data.** The third unit introduced the field of ML and its relationship to big data. The ML process was presented to the students, providing them with a holistic view of the ML pipeline. Furthermore, they developed an understanding of how ML techniques leverage vast amounts of data to extract meaningful patterns, make accurate predictions, and derive valuable insights. By comprehending the connection between ML and big data, students were expected to be able to recognize the pivotal role that ML plays in contemporary data-driven AI systems.

In the live session, students were introduced to two case studies that elucidated the application of data and ML within the health domain. The first case study involved the utilization of medical records for predicting malaria and related diseases, while the second case study demonstrated the use of medical images, specifically X-rays, for diagnosing and monitoring neuro-diseases. Then, students engaged in a discussion about the required data for these systems, as well as a critical examination of the advantages and disadvantages associated with employing these ML systems.

**Computer Vision.** The fourth unit focused on CV, explaining its general objectives and highlighting various specific CV tasks. Students gained the ability to recognize the presence of face recognition in everyday applications they use and to critically evaluate its potential benefits and risks. Also, it fostered an awareness of the potential social biases inherent in CV applications.

In the live session, a tool was used to prompt discussion on human-produced annotations for images of people. The discussion focused on the influence of annotators' backgrounds on the data, how these data are often used to train ML algorithms and how this can impact the behaviour of ML algorithms. Finally, another tool demonstrated how three CV algorithms describe images of people, prompting discussion of the generated data.

**Natural Language Processing.** The fifth unit provided students with an overview of the key applications of NLP, encompassing speech recognition, text-to-speech conversion, machine translation, and information retrieval. Students explored the capabilities of these NLP technologies, recognizing their ubiquitous presence in applications they use in their daily lives. Finally, the advantages, as well as the risks and challenges of NLP applications, were discussed.

During the hands-on activity, students reflected on their personal experiences with NLP applications. They were asked to answer the question, "What do you like and dislike about the NLP app you use?". Students actively participated by contributing their responses to a shared document, where they listed the NLP systems they use and shared both the positive and negative aspects of those systems.

**Personalization.** The sixth unit focused on the concept of personalization in AI applications, highlighting its two types: adaptable and adaptive systems. Students became aware of how AI technologies can tailor experiences to individual users, and adapt to their preferences using examples of personalization in some "everyday" applications that they use. Students were able to evaluate the benefits and potential drawbacks of personalization in AI applications.

During the live session, students discussed the presence of personalization in their daily use of AI applications. Two case studies were presented, illustrating the limitations of personalization in Google image search results. The discussion focused on instances where the algorithm assumed gender-specific roles, such as associating "female nurses" and "male surgeons" in response to user queries for keywords like 'nurse' and 'surgeon'.

**Ethical Issues.** In the seventh unit, students explored the core ethical issues of AI applications. A definition of ethics that was provided to the students stimulated discussions on how ethical principles should be applied to AI systems. Moreover, the concept of Trustworthy AI was introduced, highlighting the need for developing AI systems that consider transparency, fairness, and accountability. Lastly, the significance of human oversight in ensuring ethical AI deployment was explored.

In the live session, a real-time polling tool was employed to gather user responses. Firstly, students were prompted to share AI applications they both use and trust. Then, they were asked to articulate the essential features of AI applications to be considered reliable. A practical exercise followed, presenting a scenario involving an AI application used to shortlist resumes for an HR manager. Students were to: *a)* give possible causes of unfairness in this system, *b)* in case the system behaves unfairly who should be held accountable and *c)* whether they would trust this system.

**Living with AI.** The final unit brought everything together and emphasized the prospects and challenges associated with AI, underlining strategies for establishing a healthy symbiotic relationship with AI systems broadly. Students developed an appreciation for the importance of AI literacy and its role in fostering effective use and engagement with AI technologies. Additionally, they examined the challenges, potential dangers, and long-term benefits that arise from coexisting with AI in various aspects of our lives.

In the final session, students used again the polling tool to provide their respective definitions of AI. Then, they engaged in a discussion concerning the promotion of AI literacy in society. Finally, they delved into an examination of potential risks and challenges associated with the integration of AI into our daily lives.

### 2.4 Assignments

**Assignment 1.** Train a machine learning algorithm.

The first assignment provided students with hands-on experience in training an ML algorithm. The assignment required students to utilize the online platform "Machine Learning for Kids"[3] to train their own ML algorithm. During the assignment, students were asked to test their models twice and compare the results. The first time, they trained the model with only five examples. The second time, they trained with twenty examples and tested the same sentences. They were prompted to reflect on the differences in predictions and confidence scores provided by their model. Specifically, on one hand, when the model was trained with only five examples, it had a limited dataset to learn from, resulting in less accurate predictions. On the other hand, training the model with twenty examples provided a larger and more diverse dataset, leading to more accurate predictions with higher confidence levels.

---
[3] https://machinelearningforkids.co.uk/

The assignment highlighted the importance of data quantity, diversity, and quality in training ML algorithms. It emphasized that using an insufficient amount of training data can hinder model's ability to learn effectively. This assignment helped students to gain a better understanding of the crucial role that training data plays in the development of robust and reliable ML algorithms.

**Assignment 2.** How do Computer Vision see your photos?

The second assignment examined the nature of CV, focusing on how CV algorithms interpret our images. The assignment involved the utilization of three commercial CV services – Clarifai, Amazon Rekognition, and Google Vision – through the OpenTag platform [13].[4] Students were asked to upload a photo that they would include in a job application if they knew that a CV algorithm would be used in the process of shortlisting candidates for an interview.

Through this assignment, students gained a deeper understanding of the capabilities and limitations of CV. It provided students with a hands-on exploration of CV algorithms, enabling them to witness how these algorithms "see" and interpret visual data. It helped them understand CV technology but also, highlighted the importance of considering algorithmic outputs critically, appreciating their strengths and limitations when applying CV in real-world.

**Assignment 3.** Create a smart assistant for a smart classroom.

The third assignment involved the creation of a "smart assistant" that can control virtual devices for a smart classroom using Scratch and the Machine Learning for Kids platform.[5]. Students were given a step-by-step guide for the creation of the "smart assistant." They were asked to first try the existing "smart assistant" in Scratch, which was able to recognize only four specific commands. Then, students created and trained their ML algorithm with more examples, enabling the "smart assistant" to recognize more commands.

After developing their own smart assistant, they answered questions related to the limitations of smart assistants, the challenges faced by developers when developing that kind of technology. This assignment provided students with hands-on experience in building AI-powered smart assistants and encouraged critical thinking about the complexities and ethical considerations of NLP. The assignment was not graded and they got a pass or fail for this assignment.

**Assignment 4.** The "Filter Bubble" effect.

The fourth assignment of the AI course involved the interaction with an educational demo to explore the Filter Bubble effect within the contexts of social media and search engines [3].[6]

To interact with the demo, students had to select a pre-made profile or create their own. They then compared the results they obtained with those of other profiles, noting the differences in the information presented to them. The demo subsequently provided explanations as to why these variations occurred, shedding light on how personalization algorithms filter content based on user preferences, previous interactions, and the user profile.

The aim was to provide students with firsthand experience of how personalized algorithms can impact the information presented to users based on their profiles and browsing history. Through this assignment, they gained valuable insights on Filter Bubble effect and its implications in shaping the information users see online.

**Assignment 5.** Fairness and trust in algorithmic decision-making.

The aim of the fifth assignment was to explore students' perceptions of fairness and trust in AI decision-making. Three scenarios involving AI systems making a decision that affects humans were adapted from [5]. Students were asked to rate their agreement on six statements related to these decisions, including their agreement with the decision, understanding of the decision-making process, appropriateness of factors considered, fairness of the decision-making process, whether the individual deserved the outcome, and trust in the AI system's decision compared to a human's decision.

The objective was to assess whether students considered the decisions made by AI systems to be fair and whether they would place more trust in the decision-making capabilities of AI compared to humans. This assignment allowed students to critically evaluate the role of AI in decision-making processes and reflect on the ethical considerations surrounding the use of AI systems in various contexts. Through this assignment, students had a better understanding of the complexity of fairness and trust in AI systems.

### 2.5 Quizzes

A self-assessment quiz was available on the website after each live session. Successful completion of the course did not mandate completion of these quizzes; however, students were strongly encouraged to engage with all of them. The quizzes included 10 multiple-choice questions and were designed to provide participants with feedback regarding their understanding of each unit's content.

In contrast, the final quiz was a prerequisite for successfully completing the entire course. This conclusive quiz became available on the course website after the last live session, and students were provided with a deadline for its completion. They had the flexibility to undertake the quiz at their convenience within the allotted timeframe. The final quiz comprised 10 multiple-choice questions and 5 open-ended questions. After the deadline, one of the instructors evaluated the open-ended responses submitted by the students. Upon completion of the grading process for all assignments and the final quiz, certificates of attendance were automatically issued.

## 3 METHODOLOGY

### 3.1 Participants

Before launching the course, we applied for and received ethics approval from the University of [redacted], for any data collection, analysis and storage. A total of 40 students showed interest in the course, with 29 officially registering to the course. Of the registered students, 24 completed the required assignments and the final Quiz and successfully passed the course. The majority of the students who completed the course were males (62.5%), between 20-24 years old (66.7%). 20.8% were between 25-29, and the rest were over 30. Most students were from the Economics and Management department (20.8%), followed by the CS department (16.7%). We also had students from the Sociology (12.5%), Psychology (12.5%), and Law (12.5%) departments. The remaining students came from International Studies (8%,) Civil and Mechanical Engineering (8.3%), Biology (4.2%), and Industrial Engineering (4.2%). More information about the demographics of the students who were interested and registered to the course can be found in Table 1.

---

[4]http://ec2-3-249-45-201.eu-west-1.compute.amazonaws.com/opentag-protea/
[5]https://machinelearningforkids.co.uk/
[6]http://ec2-54-216-244-107.eu-west-1.compute.amazonaws.com/cyclops/

Table 1: Demographics.

|  | Interested | Registered | Completed |
|---|---|---|---|
| **Students** | 40 | 29 | 24 |
| **Gender** | | | |
| Male | 60% | 58.6% | 62.5% |
| Female | 40% | 41.4% | 37.5% |
| **Age** | | | |
| 20-24 | 65.0% | 65.6% | 66.7% |
| 25-29 | 27.5% | 24.1% | 20.8% |
| Over 30 | 7.5% | 10.3% | 12.5% |
| **Field of Studies** | | | |
| Law | 17.5% | 10.3% | 12.5% |
| Computer Science | 15.0% | 20.7% | 16.7% |
| Economics | 15.0% | 20.7%% | 20.8% |
| Sociology | 10.0% | 10.3% | 12.5% |
| Psychology | 10.0% | 13.8% | 12.5% |
| Engineering | 7.5% | 6.9% | 8.3% |
| Industrial Engineering | 5.0% | 3.4% | 4.2% |
| International Studies | 5.0% | 6.9% | 8.3% |
| Biology | 5.0% | 6.9% | 4.2% |
| Philosophy | 2.5% | – | – |
| Did not say | 7.5% | – | – |

## 3.2 Data Collection

At the end of the course, students were asked to complete two online questionnaires to evaluate the course and the knowledge they gained from the course. The first questionnaire (course evaluation) was completed by the students during the last live session of the course. The second questionnaire was sent out by email to all registered students – even to those who did not complete the course – one month after the completion of the course. It evaluated the knowledge students gained from the course.

A total of 20 students completed the first questionnaire. A total of 29 students completed the second questionnaire; of those, 24 completed the course, and five dropped out.

## 4 COURSE EVALUATION AND STUDENT PERFORMANCE

Half of the respondents (50%) reported that they attended all the synchronous sessions, while 10% indicated that they did not attend any "live" session, choosing instead to watch the recorded video lectures on their own time (i.e., asynchronously). In contrast, 40% reported that they attended some live sessions and watched some via the recorded video. Furthermore, 54% of respondents watched all pre-recorded video lectures before attending the live session, while half of the respondents noted that they watched the pre-recorded video lectures "sometimes." Only 5% of respondents indicated that they did not watch the pre-recorded video lectures at all.

The majority of the participants (85%) selected the higher-end of the scale (4 and 5 in a five-point Likert item) stating that the pre-recorded videos were helpful. 90% of them agreed that the teaching approach followed by the instructors was effective. They also agreed (95%) that the educational content and course materials were understandable and accessible. All of the participants indicated that the structure of the course was clear and effective. They also agreed (90%) that many examples and explanations were given for a better understanding of the material and that the assignments of the course "helped me to better understand the material" (85%). The vast majority of the respondents (85%) agreed that the course met their expectations. Similarly, the vast majority of the participants (90%) reported that the course is potentially "useful for everyone."

After completing the course, students felt that they were able to better understand behaviors (96.8%) and analyze functions of "everyday" AI applications (87.5%). They reported that they were able to recognize that methods such as ML are used for the development of AI systems (91.7%). The vast majority (91.7%) indicated that they were able to identify the use of AI technologies such as (Computer vision, NLP, and Personalization) in their daily lives. They also felt that they were able to understand the potential dangers of AI (83.3%), while 79.2% noted that they felt able to explain the ethical and social issues that can arise from the use of AI. Finally, 91.6% of them noted that they were able to understand the importance of AI literacy. The average score of the assignments and the quizzes show that learners were able to understand the topics (Table 2).

**Drop-out.** We noticed that 5 out of the 29 students registered for the course dropped out. Students were asked to complete an online questionnaire with six 5-point Likert-Scale statements aiming to understand the reason they dropped out, and one free-text question asking them to specify why they did not complete the course. Students agreed that the course being taught in English or the satisfaction with the instructors, were not the reasons for dropping out. Four of them stated that they did not complete the course because they were very busy that semester and did not have the time to commit, while two of them indicated that they did not have time to complete the assignments. One student stated that the course required more time than expected. One of them indicated that he/she did not complete the course because the course did not satisfy their expectations. Four of the students that dropped out noted that "[the course] required more time to finish, which [they] didn't have at that time" (student – s15), even though they may have "liked the course a lot" (s23). One of the students noted that he/she did not complete the course because of privacy concerns while using the teaching platform of the course.

## 5 REFLECTIONS

Next, we reflect on our students' experiences, as well as our own experience teaching the course. By summarizing these reflections, we aim to inform further the development of the course and help others who might want to pursue a similar endeavour.

**Students' Experience.** Our outcomes pointed out the impact of the course on students' perspectives and competencies. By successfully addressing both positive and negative aspects of AI, students demonstrated a comprehensive understanding of AI's potential benefits and drawbacks. Our students also indicated that they were able to analyze and explain the basic functionalities of common AI application, as well as to contemplate their possible ethical and social issues, highlights the course's effectiveness in fostering a well-rounded comprehension of AI beyond technical facets.

Table 2: Student performance (#responses to the assignments (A) and self-assessment quiz (Q), mean/median score (out of 100)).

|  | Unit 1 | | Unit 2 | | Unit 3 | | Unit 4 | | Unit 5 | | Unit 6 | | Unit 7 | | Unit 8 | | FQ |
| --- | --- | --- | --- | --- | --- | --- | --- | --- | --- | --- | --- | --- | --- | --- | --- | --- | --- |
|  | Q | A | Q | A | Q | A | Q | A | Q | A | Q | A | Q | A | Q | A | |
| responses | N/A | N/A | 24 | N/A | 21 | 27 | 22 | 25 | 19 | 25 | 16 | 25 | 15 | 25 | 16 | N/A | 24 |
| score ($\mu$) | N/A | N/A | 82.9 | N/A | 72.4 | 70.3 | 65.5 | – | 82.6 | 87.1 | 90.6 | – | 84 | – | 72.5 | N/A | 80.6 |
| median | N/A | N/A | 90 | N/A | 70 | 68 | 60 | – | 80 | 89 | 90 | – | 80 | – | 75 | N/A | 79.5 |

These findings align with previous research (e.g., [6, 7, 16, 20]), which emphasizes the efficacy of relatively short educational interventions, such as seminars and short courses, in reshaping students' perceptions of AI technologies. The reported engagement patterns, where a significant portion of students actively participated in synchronous sessions and diligently consumed pre-recorded content, further emphasize the accessibility and adaptability of the course to different learning preferences. The high satisfaction rates with course elements, including structure, teaching approach, and learning materials, indicate a positive overall learning experience.

Moreover, students' enhanced ability to identify and comprehend the use of AI in their daily lives, coupled with their increased awareness of potential dangers and ethical implications, reflects the course's success in not only imparting theoretical knowledge but also in fostering critical thinking and ethical considerations. The focus on technical understanding and ethical reflection is pivotal in preparing individuals to navigate the evolving landscape of AI.

**Instructors' Experience.** We tried to make the course flexible by addressing various learning styles. By including short, pre-recorded videos, the live sessions became more organized and allowed a more focused and efficient use of interactive time. However, we faced challenges encountered in promoting open discussions during live sessions. To overcome student hesitation in participating during live discussions, we utilised tools that allowed anonymous online responses such as shared documents and live polls. The use of those tools proved to be a considerate solution, creating an inclusive and comfortable environment for student engagement.

The live sessions provided a valuable space for sharing different viewpoints and showcasing the diverse academic backgrounds of the participants. We noticed that during the discussion students from different departments brought different viewpoints. For instance, law perspectives from lawyers, technical insights from computer science students, and unique contributions from those in other fields collectively enriched each discussion. This interdisciplinary dynamic not only improved the overall learning experience but also highlighted the course's success in promoting collaboration among students from different academic fields.

The success of hands-on assignments highlighted the essential role of practical application in the learning process. Moving beyond theory, these assignments involved tasks, such as developing an ML algorithm or a smart assistant and engaging in discussions on trendy AI topics such as ChatGPT. Through the assignments, students deepen their understanding of each topic. This approach reinforced theoretical concepts and provided students with a more immersive and practical learning experience.

In summary, our experience of teaching this course highlights the pedagogical significance of a blended online learning approach. By combining pre-recorded videos, interactive sessions, and hands-on assignments, we successfully tackled issues in online learning and at the same time created a positive, and enriching educational experience for students from various backgrounds. This approach ensured a more comprehensive and engaging learning environment, fostering the academic success of interdisciplinary students.

## 6 CONCLUSION AND FUTURE PLANS

Responding to the growing demand to educate university students about AI, we have developed and offered an introductory AI course tailored for students from any discipline. In this experience report, we have presented the experiences of our students using a data collection app throughout the duration of the course. The course evaluation indicated that upon course completion, our students were more knowledgeable about AI, and had a better understanding of the potential social and ethical issues that AI can bring.

Recognizing the crucial need for AI education and observing the positive impact of our course, we are dedicated to providing an accessible and continuously improving learning experience. Our commitment extends beyond the university setting as we plan to persist in delivering this course to diverse audiences. In our future plans, we aim to reintroduce the course, making it not only available to university students but also accessible to the broader public on a semester basis. This step underscores our dedication to extending valuable AI education across different segments of society.

Furthermore, we recognize that the most valuable feedback comes directly from those experiencing the course. Hence, we are planning to foster a more interactive evaluation session with our students. Beyond routine evaluations, we aspire to encourage students to provide more detailed feedback, sharing their specific thoughts and suggestions. This will not only allow us to pinpoint areas for improvement but will also empower students to play an active role in shaping their learning environment.

It is also important to keep in mind that the field of AI is dynamic, with constant advancements and evolving trends. To ensure our course remains at the forefront of these developments, we are committed to regularly updating our content. By integrating the latest AI trends and technologies, in order to provide our students with insights into cutting-edge applications and possibilities within the field. This will equip our students with knowledge that aligns with the current state of AI, and prepare them for the challenges and opportunities that lie ahead.

## 7 ACKNOWLEDGMENTS

This project is partially funded by the EU's Horizon 2020 Research and Innovation Programme under agreement No. 739578 (RISE).